# The precision of the arithmetic mean, geometric mean and percentiles for citation data: An experimental simulation modelling approach[1]

Mike Thelwall, Statistical Cybermetrics Research Group, School of Mathematics and Computer Science, University of Wolverhampton, Wulfruna Street, Wolverhampton, UK.

When comparing the citation impact of nations, departments or other groups of researchers within individual fields, three approaches have been proposed: arithmetic means, geometric means, and percentage in the top X%. This article compares the precision of these statistics using 97 trillion experimentally simulated citation counts from 6875 sets of different parameters (although all having the same scale parameter) based upon the discretised lognormal distribution with limits from 1000 repetitions for each parameter set. The results show that the geometric mean is the most precise, closely followed by the percentage of a country's articles in the top 50% most cited articles for a field, year and document type. Thus the geometric mean citation count is recommended for future citation-based comparisons between nations. The percentage of a country's articles in the top 1% most cited is a particularly imprecise indicator and is not recommended for international comparisons based on individual fields. Moreover, whereas standard confidence interval formulae for the geometric mean appear to be accurate, confidence interval formulae are less accurate and consistent for percentile indicators. These recommendations assume that the scale parameters of the samples are the same but the choice of indicator is complex and partly conceptual if they are not.

## 1. Introduction

The European Union and some countries and regions produce periodic reports that compare their scientific performance with the world average or with appropriate comparators (EC, 2007; Elsevier, 2013; NIFU, 2014; NISTEP, 2014; NSF, 2014; Salmi, 2015). One of the points of comparison is typically (but not always: NIFU, 2014) the citation impact of the research conducted (Aksnes, Schneider, & Gunnarsson, 2012; Albarrán, Perianes-Rodríguez, & Ruiz-Castillo, 2015; King, 2004) on the basis that this is a likely pointer to its average scientific quality or influence. Citation data is often reported in conjunction with a range of other indicators, such as expenditure, publication volumes, patenting and PhD completions. Monitoring such data over time may give insights into the success of a science system and perhaps also of individual large scale policy initiatives or restructuring. Sets of departments within a field are also sometimes evaluated with the aid of quantitative data, and other groups of researchers may also be compared for theoretical reasons, such as to contrast the impacts of collaborative and non-collaborative research (e.g., Abramo, & D'Angelo, 2015b). Whilst Mendeley readership counts have been proposed as an alternative to citation counts for articles published in recent years (Fairclough & Thelwall, 2015a), they have not yet been used in practice and disciplinary differences (Haustein, Larivière, Thelwall, Amyot, & Peters, 2014) make them unsuitable for some fields.

The typical statistic used for comparing citation impact is some form of field-normalised citation count, such as the new crown indicator, or Mean Normalized Citation Score (MNCS) (Waltman, van Eck, van Leeuwen, Visser, & van Raan, 2011a,b). This approach has problems with robustness and interpretation (Leydesdorff, & Opthof, 2011) but is still used for the convenience with which sets of publications can be compared. At the level of an individual field and year, this indicator is equivalent (other than a common scalar multiple) to the arithmetic mean of the citation counts of the articles from that field and year. For articles from multiple fields, the arithmetic mean is calculated only after field normalisation by dividing each article by the average citation count for its field, document type and year. The arithmetic mean is not ideal, however, due to the skewed nature of citation data (de Solla Price, 1976). The median (Rousseau, 2005) is suitable for skewed data but is probably to crude to be useful in many contexts. The geometric mean (Zitt, 2012) is also appropriate for skewed data and is fine grained enough for comparisons. Percentile ranks are an alternative to direct citation counting, however (Schreiber, 2013; Schubert & Braun, 1996; Tijssen, Visser, & van Leeuwen, 2002), as well as for individuals and research groups, and when multiple indicators are needed (Bornmann, Leydesdorff, & Mutz, 2013). For comparing the citation impact of countries, the proportion of a nation's share of the world's top X% of articles can be calculated. If this share is higher than X% then the nation is above the world average for the calculation. Different values of X suggest different interpretations of the results. For example, if X=50 then the percentile statistic corresponds to the nation's share of above average research, whereas if X=1 then the percentile statistic corresponds to the nation's share of the world's very high impact research.

Given the choice of (appropriately field/year/document type normalised) arithmetic means, geometric means and percentiles for citation impact comparisons, the latter two are preferable on the grounds of the skewness of citation data. Nevertheless, it is not clear whether one of these is better than the other, and whether there are theoretical grounds to prefer one particular percentile limit. In the absence of specific policy requirements or a need to report multiple statistics, a logical way to select an indicator is to choose the one that is best able to distinguish between different countries. This would mean that the best indicator is the one that is the most precise relative to the spread of likely values for different countries. Whilst the arithmetic mean should perform poorly in this regard, a previous study with empirical data found that the geometric mean was more precise than the percentage of a country's articles in the top 10% most cited (Fairclough & Thelwall, 2015), but it did not check that this was universally true and did not check other percentiles (e.g., 50%, 1%). This article addresses this issue using a different approach, experimental simulation modelling, by comparing the relative precision of the arithmetic mean, geometric mean and percentiles with a range of different parameters.

## 2. Modelling citation distributions

If the citation counts of all articles from a single field and year are examined, they typically exhibit a strong pattern that approximates a known statistical distribution. Several different distributions have been suggested as the most suitable.

### 2.1 Alternative citation distribution models

Articles from the same subject and year seem to fit the discretised lognormal distribution reasonably well (Evans, Kaube, & Hopkins, 2012; Thelwall & Wilson, 2014a) and better than most distributions tested so far. In particular, the discretised lognormal fits at least as well as the power law in almost all cases (Brzezinski, 2015) and even for the exceptions the power law only fits the tail of citation data well (i.e., ignoring articles with few citations), which excludes its use for modelling entire citation distributions.

Count data distributions are a more natural choice for citation counts because they directly model discrete data. Of these, the negative binomial distribution (Hilbe, 2011), or zero inflated, truncated or hurdle variants (Chen, 2012; Didegah & Thelwall, 2013), seem to fit citation data better than most alternatives tried, but the discretised lognormal fits citation data better than the negative binomial (Low, Wilson, & Thelwall, 2015) probably because the negative binomial does not model the very high values well. In other words, a heavy tailed distribution (Clauset, Shalizi, & Newman, 2009), such as the lognormal or power law, is needed to account for a small number of very high citation counts.

There is some evidence of a modified negative binomial stopped sum distribution fitting slightly better than the lognormal in some cases but this is impractical for use in citation analysis because of the difficulty in accurately estimating the distribution parameters (Low, Wilson, & Thelwall, 2015). The hooked (or shifted) power law also fits citation data approximately as well as the discretised lognormal (Eom & Fortunato, 2011; Thelwall & Wilson, 2014a) but also has problems with inaccuracy of parameter estimation (Thelwall & Wilson, 2014b). Whilst parameter estimation is not directly needed in the modelling here, this difficulty suggests that it would be difficult to accurately model distributions for a predefined mean and standard deviation, as needed here.

Another exception is the Yule distribution, which is for discrete data and has been shown to fit citation data approximately as well as the discretised lognormal overall and slightly better for some sets of articles, although only after excluding articles with few citations (Brzezinski, 2015). For the current article, the option of excluding uncited articles or articles with few citations from the model, as the majority of published citation modelling papers have done, is not possible because it would invalidate the statistical analyses conducted. The Yule distribution is defined only on integers greater than zero, has a mode of 1 and is strictly decreasing. An offset of 1 could be used to allow it to model zeros but it cannot accurately model distributions that are not strictly decreasing, such as many fields with high citation counts. From the author's previously-analysed data, this includes the Scopus category of Catalysis from 2009 [mode 6] as well as the 2004 Scopus categories of Tourism [mode 4], and Developmental Biology [mode 9]. Moreover, 22 of the 45 Scopus medical subject categories from 2009 analysed in one article had a mode of at least 2, and 6 additional fields had a mode of 1 (author's re-analysis of data from: Thelwall & Wilson, in press). Related to this, the Yule distribution is probably unable to model the hook shape at the top left hand corner of most citation distributions. Hence the Yule distribution is inadequate for the complete range of citation data.

The discretised lognormal alone is used here on the basis that the Yule distribution is not suitable, the hooked/shifted power law and stopped sum have parameter estimation issues, and there would be little advantage in repeating all experiments with additional, similar distributions if there are uncertainties about their applicability.

## 2.2 The discretised lognormal distribution

The discretised lognormal is derived from the continuous lognormal distribution $e^{\mu + \sigma Z}$, where $Z$ follows the standard normal distribution with mean zero and standard deviation 1. Equivalently, if $X$ is a lognormal variable then $ln(X)$ is normally distributed. The probability density function for the continuous lognormal distribution is

$$f(x) = \frac{1}{x\sigma\sqrt{2\pi}} e^{-\frac{(\ln(x)-\mu)^2}{2\sigma^2}} \tag{1}$$

The probability of a value of $x$ occurring in the range $[a, b]$ is therefore given by the integral $\int_a^b f(x)dx$ where $x$ must be strictly positive (Limpert, Stahel, & Abbt, 2001). The two parameters for the lognormal distribution are the mean (or location) parameter μ and the scale parameter σ.

For the discretised lognormal distribution, the probability density function $f(x)$ with a scaling adjustment can be used as a point mass function, so that the probability of a value $x$ is given by $Af(x)$, where $A = 1/\sum_{x=1}^{\infty} f(x)$ is chosen to make the sum of all probabilities equal to unity: $\sum_{x=1}^{\infty} Af(x) = 1$. Alternatively, the probability of a discrete value x could be calculated as by the integral of the unit interval around $x$, $\frac{1}{B}\int_{x-0.5}^{x+0.5} f(x)dx$, where $B = \int_{0.5}^{\infty} f(x)dx$ compensates for the missing interval (0,0.5] that does not correspond to any integer value because zeros are excluded. These two versions of the point mass function for the discretised lognormal distribution are similar but not identical, particularly for values of $x$ close to the location parameter. The second version of the point mass function is used here (for specific details, see: Gillespie, 2013).

Holding the scale parameter constant, increasing the mean parameter increases the expected mean of any sample generated from the (continuous) lognormal distribution (although the mean parameter is not the same as the expected mean, which is $e^{\mu+\sigma^2/2}$ – intuitively, increasing the scale parameter increases the mean because it increases large values, which are not bounded above, more than it decreases small values, which are bounded below by 0) and so the research goal of this paper can be operationalised as testing which indicator is best able to distinguish between random samples generated from discretised lognormal distributions with the same scale parameter but different mean parameters.

A limitation of this approach is that it ignores uncited articles since $x > 0$, but this can be resolved by either ignoring all uncited articles or by adding 1 to all citation counts (i.e., including both cited and uncited articles so that all counts are shifted by 1). The latter approach is followed here. Although the geometric mean is a good estimator for the mean parameter (but not the mean) of the continuous lognormal distribution (Limpert, Stahel & Abbt, 2001), this relationship is broken by the addition of 1 and the discretisation process.

## 3. Research questions

This article uses experimental simulation modelling to assess the precision of different national citation impact indicators for individual fields, document types and years. The focus is restricted to a single field, document type and year in order to assume a relatively homogenous citation distribution. Any mixing of different fields, document types or years would greatly complicate comparisons because of the extra variables introduced. As argued above, it is important to identify the most effective indicator to distinguish between the citation impacts of different sets of articles. This study is therefore primarily driven by the following research question.

- Which of the arithmetic mean, geometric mean, and proportion in the top X% are the best able to distinguish between citation distributions with different average citation counts?

Even if, in theory, two distributions can be distinguished by a metric, in practice, evidence about the significance of any differences is dependent on the accuracy of the confidence interval formula used for the metric. A secondary goal is therefore to assess the accuracy of standard confidence interval formulae for the indicators from the perspective of modelling. One previous study has questioned their accuracy (Fairclough & Thelwall, 2015b) but, if this is incorrect then they would be useful to decide whether differences between countries were statistically significant. The second research question addresses this issue.

- How accurate are standard formulae for confidence intervals for the geometric mean and proportion in the top X% most cited for data from citation distributions?

# 4. Methods

A series of experiments were run to detect which indicator is best able to differentiate between countries having different average (mean) rates of attracting citations to their work within a single field and year. For each test, citation distributions were modelled as discretised lognormal (Thelwall, & Wilson, 2014), assuming relatively low differences in citation rates within subfields. For simplicity, world citations were modelled using three separate distributions: one for country 1, one for country 2, and one for the rest of the world. When lognormal distributions with different location parameters are combined then the resulting distribution is not necessarily lognormal. The case is analogous to the situation with the normal distribution: combining two normal distributions with very different means can generate a bimodal distribution. Nevertheless, as long as the distribution means are not too different then their combination should be approximately normal and the same is true for the lognormal distribution. This approximation is a limitation of the method used here, however.

Each of the three distributions (country 1, country 2, and the rest of the world) was given the same scale parameter but their location parameters were allowed to vary. For the models, the extent to which the country distributions differ from the rest of the world is ignored in order to answer the first research question by detecting differences between the two countries. In other words, the first research question is interpreted as assessing the ability to distinguish between two specific countries within an international context. For simplicity, the overall mean and standard deviation parameters were set at (approximately) 1 using the methods described below. Even though these can vary and will affect the modelling results, the choice of a single value should not affect the overall pattern of the findings.

## 4.1 Parameters varying between tests

To give a range of differences between two countries, the mean parameters were varied independently from 0.9 to 1.1 in steps of 0.02 for each country. These values were chosen to capture the region of uncertainty where the differences are not large enough that all methods would be able to identify them. The two countries were allowed to have proportional shares of the world's articles of 0.05, 0.1, 0.15, 0.2 and 0.25, independently. This excludes medium sized countries with weak science bases and small countries for which large differences would be needed to appear as statistically significant. The models were constructed for sample sizes of 500, 1000, 5000, 10000, and 50,000 to cover up to the largest feasible collection of articles from a single field and year. For example, for one of the tests, country A had a mean parameter of 0.9, country B had a mean parameter of 0.92, country A had 0.05 of the world's articles, country B had 0.2 of the world's articles and the overall sample size (i.e., the total number of articles published in the selected field and year) was 5000. Each of the numbers in the previous sentence was varied independently of all the other numbers in the sentence. In other words:

- Country 1 has articles that are randomly drawn from the discretised lognormal distribution with a mean parameter of $\mu_1$, a standard deviation parameter of 1, and a proportion $p_1$ of the world's $N$ articles from a specific subject and year.
- Country 2 has articles that are randomly drawn from the discretised lognormal distribution with a mean parameter of $\mu_2$, a standard deviation parameter of 1, and a proportion $p_2$ of the world's $N$ articles from a specific subject and year.

The free parameters are subject to the conditions:

- $\mu_1, \mu_2 \in \{0.9, 0.92, ..1.1\}$, with $\mu_1 < \mu_2$ since the test is symmetric (so the cases $\mu_1 > \mu_2$ are redundant) and if $\mu_1 = \mu_2$ then there is no difference to test for. This gives 55 variations.
- $p_1, p_2 \in \{0.05, 0.1, 0.15, 0.2, 0.25\}$ independently. This gives 5×5=25 variations.
- $N \in \{500, 1000, 5000, 10000, 50000\}$. This gives 5 variations.

Since all the variations are independent, the total number of parameter sets tested is 55×25×5=1375×5=6875. These variations are all listed in the online spreadsheet associated with this paper (http://dx.doi.org/10.6084/m9.figshare.1617835), with a separate worksheet for each value of $N$. The online spreadsheet includes additional cases for $\mu_1 = \mu_2$ that were processed as a testing check.

The (continuous) lognormal distribution with location parameter $\mu_0$ and scale parameter $\sigma$ has mean $m = e^{\mu_0 + \sigma^2/2}$. If proportions $p_1$ and $p_2$ of a sample are replaced with distributions with the same scale parameters but location parameters $\mu_1$ and $\mu_2$, respectively, then the mean of the overall distribution will be:

$$p_1 e^{\mu_1 + \sigma^2/2} + p_2 e^{\mu_2 + \sigma^2/2} + (1 - p_1 - p_2) e^{\mu_0 + \frac{\sigma^2}{2}}. \tag{2}$$

After fixing $\mu_1$ and $\mu_2$, the location parameter $\mu_0$ can also be adjusted to ensure that the overall distribution mean is equal to $e^{\mu + \sigma^2/2}$. Solving

$$e^{\mu + \sigma^2/2} = p_1 e^{\mu_1 + \sigma^2/2} + p_2 e^{\mu_2 + \sigma^2/2} + (1 - p_1 - p_2) e^{\mu_0 + \sigma^2/2}$$

for $\mu_0$ gives:

$$\mu_0 = \ln\left(\frac{e^\mu - p_1 e^{\mu_1} - p_2 e^{\mu_2}}{1 - p_1 - p_2}\right). \tag{3}$$

This formula was used to keep the arithmetic mean constant at $e^{\mu + \sigma^2/2}$ for the overall citation distribution as the proportions and means of the subsamples changed. The geometric mean for the overall population may change, however. Although the standard deviation is 1 for each of the three samples (country 1, country 2 and the rest of the world), because they may have different means the overall standard deviation may be slightly different from 1.

The models were experimentally simulated 1000 times for each of the 6875 sets of parameters. Considering each value of $N$ separately, the raw data therefore consists of 1000×1375×(500+1000+5000+10000+50000)=91,437,500,000 integers randomly sampled from the discretised lognormal distribution, each representing the citation count for a single article. Conceptually, for each of the 6875 parameter sets this can be thought of as assuming that two countries have unchanged science systems for 1000 years, randomly generating their articles for each of these years using the same set of unchanged parameters. Although the comparisons focus on differences between the two countries (which together account for 30% of the data), the rest of the world citation counts are needed for the top X% indicators and so all 91 trillion citation counts were randomly generated.

The citation counts were generated by submitting the relevant set of parameters to the R discretised lognormal generating function dislnorm inside the poweRlaw package (Gillespie, 2013, 2015).

## 4.2 Indicator calculations

The arithmetic mean and geometric mean use standard formulae. For the geometric mean, 1 is added to the citation counts before calculating the geometric mean and then 1 is subtracted from the result. This shift of 1 is a standard method of allowing the geometric mean to include uncited articles. See http://altmetrics.blogspot.co.uk/2015/10/geometric-means-for-citation-counts-and.html for a worked example of the geometric mean with an offset of 1.

There are different ways of calculating percentile statistics when there are rank ties at the cut-off point. The most appropriate, and the one used here, is to allocate a proportional share of the higher rank to articles on the threshold (Waltman, & Schreiber, 2013). For example, if the total number of articles analysed is 100 and the cut-off for the top 1% is 10 citations but three articles have 10 citations and none have more, then each article would count as 1/3 of an article for top 1% calculations.

## 4.3 Experimental simulation confidence intervals

From the 1000 iterations of each parameter set it is possible to check how substantial the differences between citation indicators are for the two countries. Using the Null Hypothesis Significance Testing (NHST) philosophy, a standard approach to compare parameters of the experimentally simulated samples would be to test the null hypothesis that the two samples were drawn from populations with the same population mean against the alternative hypothesis that they were drawn from populations with different population means. The test that most often correctly rejected the null hypothesis would therefore be the most statistically powerful and the best to use in practice. As argued in the Appendix, however, existing standard statistical tests for differences between the modelled countries in terms of proportion of articles in the top X% are inappropriate because they are sensitive to the number of articles in a sample and so a non-standard confidence interval approach was used instead.

Even though standard statistical tests cannot be used for the differences between two samples of the types generated here, it is still possible to calculate confidence intervals for each individual sample. Non-overlapping confidence intervals suggests that differences between the two countries could be reliably detected with a metric. For real citation data, however, all statistical tests and confidence intervals might be problematic if they do not consider the possible dependency of measurements (Schneider, 2013), and should be interpreted cautiously.

For each of the 6875 parameter sets, the results of the 1000 iterations of each model were used to calculate 95% confidence intervals for the model distribution mean, geometric mean, and proportion in the top X cited articles, where X=1%, 10%, and 50%, for each of the two countries. This gave a total of 6875×5×2=68750 confidence intervals. Each confidence interval was calculated by arranging the relevant statistic for the 1000 iterations by size and selecting the 25th smallest for the lower confidence limit and the 25th largest for the upper confidence limit. This ensures that at least 95% of the modelled statistics from the 1000 iterations fell within the confidence intervals. For example, the 95% confidence interval for the sample arithmetic mean for the first country for each of the 6875 parameter sets contained at least 95% of the 1000 modelled sample means for that parameter set.

## 4.4 Confidence interval formulae

In addition to the experimental simulation confidence intervals, standard formulae were used to calculate confidence intervals for the mean. For a geometric mean from a lognormal distribution, since the natural log of the discretised lognormal distribution is approximately normal (the natural log of the continuous lognormal distribution is exactly normal, but the discretisation process breaks this relationship), it seems reasonable to use the standard formula for sample confidence limits on the log transformed data:

$$\bar{x} \pm t_{\alpha/2,n-1}s/\sqrt{n}, \tag{4}$$

Here, $\bar{x}$ is the sample mean, $t_{\alpha/2,n-1}$ is the appropriate figure from the t distribution, $s$ is the sample standard deviation and $n$ is the sample size. For large samples and 95% confidence intervals, as used here, $t_{\alpha/2,n-1} = 1.96$, which is the normal distribution value. When applied to citation counts, $c$, this formula can be applied to $ln(1+c)$ and then transformed back with $exp(y) - 1$ so that the limits are

$$\exp\left(\overline{ln(1+c)} \pm \frac{t_{.05,n-1}s}{\sqrt{n}}\right) - 1 \tag{5}$$

Here, $s$ is the standard deviation for $ln(1+c)$, which should be approximately the scale parameter of the lognormal distribution. For simplicity, however, the figures reported are for the logarithmically scaled data $ln(1+c)$ before the transformation back with the exponential function. All of these formulae are for continuous data and will therefore be approximations for discrete data. Moreover, the addition of 1 for the geometric mean calculations makes the geometric mean formula a further approximation.

For the arithmetic mean of the lognormal distribution there is not an exact confidence interval formula (Land, 1972; Zhou & Gao, 1997) but a bootstrapping approach (i.e., in the context of this paper, resampling from the modelled data for each iteration) could be used instead to estimate confidence limits – although Cox's method is an appropriate alternative for large sample sizes (Land, 1972; Zhou & Gao, 1997). Confidence intervals with the bootstrapping approach were not calculated, however, because it seems misleading to apply bootstrapping to data that is already simulated and may be from sample sizes as small as 25 for individual countries.

For the proportion in the top X% indicator, the normal approximation to the binomial can be used to generate a confidence interval formula although, as argued in the Appendix, it is not reliable in some cases. If the sample proportion in the top X% is $p$ out of a data set of $n$ papers then, as long as the population proportion is not too close to 0 (or 1) and the sample size is large enough (which is often not the case for the top 1% indicator), the sample proportion $p$ is approximately normally distributed (De Moivre, 1756) and a confidence interval can be estimated with the standard formula

$$p \pm z_{\alpha/2}\sqrt{p(1-p)/n} \tag{6}$$

where $z_{\alpha/2} = 1.96$ for a 95% confidence limit (Box, Hunter, & Hunter, 1978). In this situation, the binomial formula is itself an approximation because of the presence of ties in rankings – which are dealt with using the Waltman-Schreiber approach (Waltman & Schreiber, 2013).

## 4.5 The confidence interval similarity formula

Because of the necessity to avoid hypothesis tests (see Section 4.3) and the large numbers of confidence intervals to be compared at different parameter values (6875 pairs of two confidence intervals for each indicator), an alternative method was needed to assess the statistical significance of the difference between two sample means in terms of the extent to which their confidence intervals overlap, or their distance apart when they did not overlap. This would be similar to the *p* value in a hypothesis test but focusing instead on the confidence intervals. Given two modelled sample means $\overline{x_1} < \overline{x_2}$, with modelled confidence intervals $[x_{1L}, x_{1U}]$ and $[x_{2L}, x_{2U}]$, respectively, it would be possible to test whether each mean was in the other confidence interval. These checks of $\overline{x_2} \in [x_{1L}, x_{1U}]$ and $\overline{x_1} \in [x_{2L}, x_{2U}]$ are not equivalent although they probably give the same results in almost all cases. These tests can also be misleading because in some cases, as described below, the mean of a sample is not in its own 95% confidence interval.

Instead, the (half) widths of the two confidence intervals relative to the difference between two means is assessed with the following *confidence interval similarity formula*:

$$\frac{(x_{1U} - \overline{x_1}) + (\overline{x_2} - x_{2L})}{2(\overline{x_2} - \overline{x_1})} \qquad (7)$$

In the ideal boundary case where each mean is exactly on the limit of the other sample's confidence interval, $\overline{x_1} = x_{2L}$ and $\overline{x_2} = x_{1U}$ and (7) is equal to 1. If $\overline{x_1}$ is $\partial_1$ inside the second confidence interval but $\overline{x_2}$ is $\partial_2$ outside the first confidence interval then confidence interval similarity formula (7) will be less than 1 as long as $\overline{x_2}$ is further inside the second confidence interval than $\overline{x_1}$ is outside the second confidence interval: $\partial_1 \geq \partial_2$, which seems reasonable. In contrast, if both are outside of the other's confidence intervals then (normally – see the paragraph below) $\overline{x_1} < x_{2L}$ and $\overline{x_2} > x_{1U}$ and so (7) is greater than 1. Thus, when the confidence interval similarity formula is greater than or equal to one, at least one mean is likely to be inside the 95% confidence interval of the other, whereas if it is less than one then it is likely that neither are. This formula was used for the arithmetic and geometric means, as well as for the proportion of a nation's output in the top X%.

The confidence interval similarity formula has anomalous behaviour when the sample consists of mostly zeros because then each mean can be outside of its own confidence interval. For example, in a sample of 100 numbers, all of which are zero except for a single one, the arithmetic mean is 1/100 but a 95% confidence interval constructed by bootstrapping or experimental simulation would be [0,0]. This can occur with the top 1% statistics for countries that produce a small number of below average outputs and are very unlikely to have co-authored any top 1% publications. In this unusual case it is possible that $\overline{x_1}$ is outside the second confidence interval because $\overline{x_1} > x_{2U}$ rather than for the usual reason that $\overline{x_1} < x_{2L}$ even though $\overline{x_1} < \overline{x_2}$. In this case it is likely that both confidence intervals are [0,0] and the formula 7 is then equal to 0.5. A value of less than 1 in this situation seems consistent with the other values of the formula in this case since both confidence intervals are the same.

The R code used for the calculations has been placed online (http://dx.doi.org/10.6084/m9.figshare.1617835), as have the full results.

## 5. Results

The results show that the geometric mean is best overall at distinguishing between subsamples having different means (Table 1). The top 50% is almost as good, as is the arithmetic mean, but the top 1% is by far the weakest.

**Table 1**. The number of configurations of the two subsamples out of 1375 for each sample size for which the confidence interval similarity formula is less than 1 (based on 1000 repetitions of the models for each set of parameters), indicating evidence of a difference between the subsamples. Each subsample varies independently in mean parameter from 0.9 to 1.1 in steps of 0.02 and in size from 5% to 25% in steps of 5%.

| Sample size | Arithmetic mean | Geometric mean | Top 1% | Top 10% | Top 50% |
|---|---|---|---|---|---|
| 500 | 0 (0%) | 5 (0%) | 0 (0%) | 0 (0%) | 1 (0%) |
| 1000 | 18 (0%) | 83 (6%) | 0 (0%) | 4 (0%) | 58 (4%) |
| 5000 | 429 (31%) | 609 (44%) | 6 (0%) | 306 (22%) | 533 (39%) |
| 10000 | 668 (49%) | 819 (60%) | 71 (5%) | 545 (40%) | 870 (56%) |
| 50000 | 1087 (79%) | 1158 (84%) | 563 (41%) | 1018 (74%) | 1139 (83%) |

As illustrated for one case in Figure 1, the confidence interval overlap indicator tends to be lowest overall for the geometric mean, slightly higher for the top 50% and higher still for the arithmetic mean.

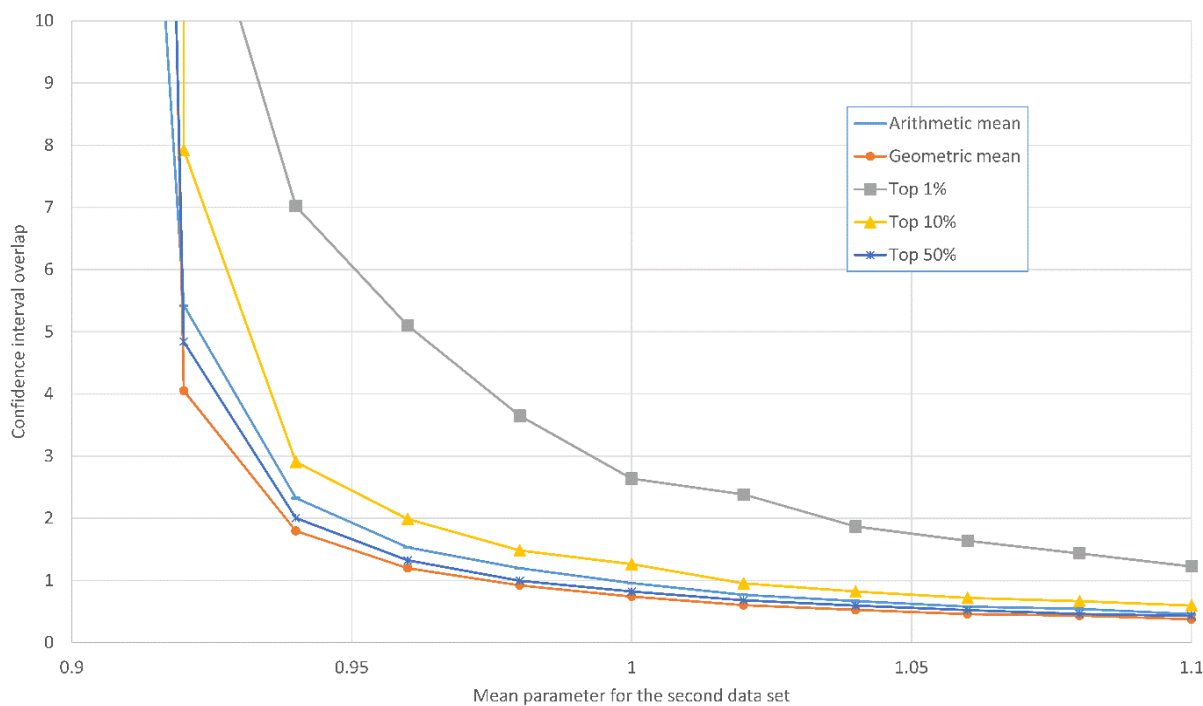

Figure 1. Confidence interval overlap figures for a sample size of 5000, two subsample sizes of 500 each, a first subsample mean parameter of 0.9, and overall mean and standard deviation (scale) parameters of 1 (based on 1000 repetitions of the models for each set of parameters). A score below 1 suggests that the means of the two subsamples are inside the other's 95% confidence interval.

The confidence intervals for the geometric mean calculated with a formula agree with the modelled confidence intervals to within 7% in all cases (Table 2). The formula results seem therefore to be precise enough to be used in practice with confidence on data that follows the discretised lognormal distribution. In contrast, the formulae for confidence intervals for the proportion in the top X% have wide discrepancies with the model results. As shown by the mean column, they tend to be conservative, meaning that they are wider than the modelled limits, but have substantial variations, as reflected in higher maximum and minimum scores as well as a higher standard deviation. The conservatism reduces the statistical power of a percentile-based statistical test, which is the ability of percentile-based tests to distinguish between nations with genuine underlying differences, because larger differences would be needed to generate non-overlapping confidence intervals. A continuity-corrected version of the formula would therefore be even less powerful. More worryingly, the substantial negative minimum values (Table 2) also undermine confidence in the results, and so the normal approximation to the binomial formula should only be used cautiously for percentile (top X%) indicators. The largest minimum values originated from the small sample sizes, for which the percentile statistics are the least precise. This is presumably because the normal distribution is a poor approximation to the binomial distribution for small sample sizes.

**Table 2**. The difference between 95% confidence interval limits derived from a formula and 95% confidence interval limits derived from the modelling (the difference between the modelling and formula limits, divided by the width of the modelling 95% confidence interval). Positive figures indicate that the formula is conservative (n=1375 for each row).

| | Sample size | Min. | Max. | Mean | Std. dev. |
|---|---|---|---|---|---|
| Geometric mean lower | 500 | -6% | 7% | 0% | 2% |
| Geometric mean upper | 500 | -6% | 9% | 1% | 2% |
| Top 1% lower | 500 | 8% | 132% | 43% | 21% |
| Top 1% upper | 500 | -65% | 126% | 1% | 27% |
| Top 10% lower | 500 | -26% | 96% | 11% | 19% |
| Top 10% upper | 500 | -36% | 95% | 5% | 20% |
| Top 50% lower | 500 | -29% | 95% | 10% | 20% |
| Top 50% upper | 500 | -29% | 94% | 10% | 19% |
| Geometric mean lower | 1000 | -6% | 7% | 0% | 2% |
| Geometric mean upper | 1000 | -6% | 10% | 1% | 2% |
| Top 1% lower | 1000 | 2% | 112% | 27% | 18% |
| Top 1% upper | 1000 | -56% | 104% | 1% | 23% |
| Top 10% lower | 1000 | -27% | 95% | 10% | 19% |
| Top 10% upper | 1000 | -35% | 97% | 6% | 20% |
| Top 50% lower | 1000 | -29% | 99% | 10% | 20% |
| Top 50% upper | 1000 | -29% | 97% | 10% | 20% |
| Geometric mean lower | 5000 | -7% | 10% | 0% | 2% |
| Geometric mean upper | 5000 | -6% | 10% | 0% | 2% |
| Top 1% lower | 5000 | -24% | 97% | 12% | 19% |
| Top 1% upper | 5000 | -38% | 93% | 4% | 20% |
| Top 10% lower | 5000 | -27% | 95% | 9% | 19% |
| Top 10% upper | 5000 | -32% | 93% | 7% | 19% |
| Top 50% lower | 5000 | -29% | 95% | 10% | 20% |
| Top 50% upper | 5000 | -29% | 93% | 10% | 20% |
| Geometric mean lower | 10000 | -7% | 10% | 0% | 2% |
| Geometric mean upper | 10000 | -6% | 8% | 0% | 2% |
| Top 1% lower | 10000 | -26% | 94% | 11% | 18% |
| Top 1% upper | 10000 | -37% | 93% | 5% | 20% |
| Top 10% lower | 10000 | -29% | 90% | 9% | 19% |
| Top 10% upper | 10000 | -32% | 94% | 7% | 19% |
| Top 50% lower | 10000 | -32% | 95% | 10% | 20% |
| Top 50% upper | 10000 | -30% | 97% | 10% | 20% |
| Geometric mean lower | 50000 | -6% | 8% | 0% | 2% |
| Geometric mean upper | 50000 | -6% | 7% | 0% | 2% |
| Top 1% lower | 50000 | -28% | 97% | 9% | 19% |
| Top 1% upper | 50000 | -32% | 92% | 6% | 20% |
| Top 10% lower | 50000 | -28% | 94% | 8% | 19% |
| Top 10% upper | 50000 | -31% | 92% | 8% | 19% |
| Top 50% lower | 50000 | -30% | 98% | 10% | 20% |
| Top 50% upper | 50000 | -29% | 100% | 10% | 20% |

# 6. Discussion

An important limitation of the findings is the underlying assumption that the citation counts for articles from individual nations can be modelled by the discretised lognormal distribution. Whilst this distribution has previously been shown to be appropriate for citation data, there may be other distributions that are more accurate, and the publications from individual nations may not follow it, especially when including (appropriately normalised) citation counts from multiple fields. Moreover, the distribution assumes that all observations are independent, whereas individual researchers or research groups in a small country may tend to produce research attracting similar numbers of citations. The statistical model assumes that each of the three groupings (country 1, country 2 and the rest of the word) are homogeneous, in practice, they are each complex aggregates of sets of researchers, departments and universities (and countries for the rest of the world set), each of which may follow different distributions. In addition, there may be different distributions for each specialism within a field.

The choice of overall mean and standard deviation in the modelling as well as the range of different subsample means compared is also a limitation because these affect the percentages of differences found (Table 1). For some types of real data, the differences found may be large enough that all indicators are sufficiently discriminative (e.g., large countries compared over all articles within a single year) or, conversely, the differences may be so small that no indicators can discriminate between them (e.g., two small countries with weak science bases). Hence, the results should be interpreted in the context of any data sets analysed. The software used for the modelling is available online free (Thelwall, 2015) and the parameters used can be varied to test different scenarios.

Holding the standard deviation (scale) parameter constant is another limitation of the study because previous research has shown that the main source of variation between country citation distributions in broad fields (i.e., splitting science into just 8 different fields) is due to the differences in percentages of very highly cited articles (Albarrán, Perianes-Rodríguez, & Ruiz-Castillo, 2015; see also the extended discussion in sections below). This difference in percentages of very highly cited articles suggests that different standard deviation parameters for countries are likely. Nevertheless, the study used very broad fields and whole citation counting (although verified for fractional citation counting in two cases: Albarrán, Perianes-Rodríguez, & Ruiz-Castillo, 2015). More investigations are needed into the highly cited articles from countries with a low average number of citations in a field to check whether they are anomalous in the sense of originating from the same small group of researchers (i.e., violating statistical independence assumptions) or from internationally collaborative research (and potentially not reflecting the power of the science base of the weak country).

A general limitation is that any indicator based upon citation counts does not measure scientific excellence but is, at best, a number that tends to reflect impact within academia. Although citation counts are poor indicators of scientific impact for individual articles or researchers because of the spurious reasons for which articles can attract, or not attract, citations (MacRoberts & MacRoberts, 1996; Seglen, 1998), it is reasonable to use this data for larger groups, such as entire nations, on the basis that these spurious reasons are likely to even out over large data sets (van Raan, 1998). Nevertheless, if researchers in a country are incentivised to attract citations, such as when indicators used to assess them are, in part, derived from the citations to their articles (e.g., Abramo, & D'Angelo, 2015b), then they may gravitate towards more highly cited areas of their fields, skewing international comparisons based upon any type of citation indicator. Hence, even when differences between nations are evident in non-overlapping confidence intervals, this evidence is not proof of underlying differences in research excellence.

The importance of precision in a citation-based indicator depends partly upon the gap in indicator values between the countries compared and partly on the number of articles published in a given year by the countries concerned. As a very approximate guideline, Table 1 suggests that 50,000 articles would be enough to differentiate between countries except for very small differences. Within Scopus, 13 countries produced at least 50,000 citable documents in 2014 (the smallest was Brazil [56368] and the Russian Federation is just underneath [49018]) (see the SCImago Journal & Country Rank lists, selecting a specific year: http://www.scimagojr.com/countryrank.php). Larger differences could be detected at smaller sample sizes. At 5000 citable documents, this would encompass 53 nations in 2014 (including Slovenia [5000], but with Nigeria [4815] just missing out). Hence, the use of more precise indicators is particularly important for detecting differences between countries with low rates of publication (e.g., all of Africa except for South Africa, most of Eastern Europe, most of Central America, most of the Middle East). This applies to overall comparisons, and implies that comparisons between countries for individual subject areas would only be able to detect substantial differences, because of the much smaller numbers of publications involved.

## 6.1 Extension to the case of multiple fields with the same scale parameter

If the results are extrapolated to indicators for multiple fields, then, except for the top X% indicators (where the top X% could be calculated separately by field), the citation counts would need to be individually normalised by field, by dividing by the international field average for the document type and year, as in the MNCS indicator. Ignoring discretisation issues, MNCS normalises by dividing by (an estimator of) the population mean and so effectively generates for each separate field a lognormal distribution with unchanged scale parameter σ and location parameter equal to $-\sigma^2/2$. In the simplest case, if one country has the same citation advantage over another country in all fields analysed (in terms of differences in the location parameter) and these fields have the same scale parameter then the MNCS reduces to the single field, single distribution case analysed in Table 1 and so the same conclusions about the relative merits of the different indicators would apply.

In practice the scale parameter is likely to vary between fields and the citation advantage of one country over another is also likely to vary between fields, and so the situation may be different for the general case. Although there has been a claim that the citation distributions for a single field and year all follow a lognormal distribution with a universal scale parameter $\sigma^2 = 1.3$ reasonably well (Radicchi, Fortunato, & Castellano, 2008), this does not have strong evidence because the 20 Web of Science categories analysed, excluding uncited articles (N varying from 266 to 9761), gave $\sigma^2$ values from 1.0 to 1.8. Another article fitting the discretised lognormal distribution to 20 Scopus subject categories, excluding uncited articles (N varying from 455 to 4811), fitted scale parameters with $\sigma^2$ between 1.0 and 2.8 (Rehabilitation, N=2904) (Thelwall & Wilson, 2014a). Even within the single broad area of medicine variations are evident. An analysis of 44 medical subject areas, excluding uncited articles (N varying from 857 to 8514 – excluding the small N=45 case), found $\sigma^2$ between 1.1 and 1.9 (Thelwall & Wilson, in press). The assumption of field normalised citation counts fitting the lognormal distribution with a constant scale parameter also contradicts empirical evidence for sets of articles for the same university (Perianes-Rodríguez, & Ruiz-Castillo, in press). The same is true for sets of articles from entire countries, which exhibit substantially different amounts of variation about the mean (see Figure 3 of: Albarrán, Perianes-Rodríguez, & Ruiz-Castillo, 2015).

In summary, although the case of multiple fields with the same scale parameter is equivalent to the single field case, after field normalisation, it is probably rare in practice.

## 6.2 Extension to the case of multiple fields with varying scale parameters

In all of the experiments the scale parameters of the two samples and the remaining articles have been fixed at 1. The situation is more complex if the scale parameters differ between the samples compared because the scale parameter influences the arithmetic mean more than it influences the geometric mean.

In the continuous case and without the addition of 1, the scale parameter does not affect the arithmetic mean of the natural log of data generated by the (continuous) lognormal distribution because the expected value of the mean of the log of the data is equal to the mean parameter $\mu$. Because of this the geometric mean is unaffected by the scale parameter $\sigma$ and is equal to $e^{\mu}$. In contrast, the arithmetic mean of continuous lognormal data has expected value $e^{\mu+\sigma^2/2}$, which increases as the scale parameter increases. Increasing the scale parameter therefore increases the arithmetic mean but not the geometric mean in the continuous case. Increasing the scale parameter also increases the number of highly cited articles and so increases the proportion of articles in the top 1% and in the top 10% but does not affect the median and hence the proportion of articles in the top 50%.

The same is not true for the discretised case: increasing the scale parameter can alter the geometric mean because the offset of 1 used before taking the natural log of the data breaks the direct relationship between the mean parameter and the sample mean of the log of the data. Moreover, it is possible (in both the continuous and discrete lognormal cases) that one sample has a higher geometric mean than the other but a lower arithmetic mean. Thus, in this situation it is a conceptual rather than statistical issue as to whether the arithmetic mean (or top 50%), or top 1% or top 10% or geometric mean should be used based upon the type of central tendency that it is best to measure for a specific application. The arithmetic mean may be preferred, for example, if extra weight is desired to be given to very highly cited articles, and the geometric mean may be preferred if the data set contains articles that cannot be adequately field normalised, for example because the field categorisation used is too broad. This is because the geometric mean is superior to the arithmetic mean for averaging data sets containing values originating from different scales (Fleming & Wallace, 1986).

Allowing the difference between a country and the international average to vary by field would result in a non-lognormal distribution for each individual country when combining normalised fields (e.g., Asmussen & Rojas-Nandayapa, 2008) and so a modelling approach with random sampling, as reported in Table 1, would be needed to assess the precision of the indicators in such a context. Given the wide variety of fields and nations, it may not be possible to provide a general solution to this unless tests for many different countries and fields all give the same conclusions. Hence, the general MNCS case is a complex combination of conceptual issues (variations in the scale parameter affecting the indicators differently) and distribution issues. It is therefore impossible to find a simple recommendation that would apply to all cases. Finally, it may be more appropriate to use the geometric mean to field normalise the MNCS, as previously suggested (Lundberg, 2007; see also: Thelwall & Sud, in press).

## *6.3 Extension to the case of the same field in different years*

If sets of articles from the same field but different years are to be analysed, such as to compare monodisciplinary departments within a discipline, then it seems likely that the scale parameters would be similar for the field in different years. If testing proves this to be true, then, after MNCS-type normalisation for the different years, the main conclusions of this article about a single field and year would also apply. In other words, it seems likely that the geometric mean is the preferable indicator for normalised citation count data for a single field over multiple years.

## 7. Conclusions

The results of the citation modelling based upon the discretised lognormal distribution show that the geometric mean is the most precise, the proportion of a nation's articles in the top 50% is almost as precise and that the proportion of a nation's articles in the top 1% is by far the least precise, at least under the assumption that the countries compared and the overall sample have the same standard deviation parameter. The geometric mean is therefore the recommended citation impact statistic for comparing nations for a single field, on the assumption of equal scale parameters. The top 50% indicator is almost as good and has the advantage of being more intuitively straightforward than the geometric mean and hence may be more palatable to high level policy makers. The less precise indicators may also be used if it is desirable to present a range of indicators, especially for countries that produce enough articles to allow the indicators to be reasonably precise. These conclusions also apply if alternative citation like data, such as Mendeley readers (Fairclough & Thelwall, 2015b), are used for comparisons between sets of articles.

The results also confirm that the standard confidence interval formula for the geometric mean is accurate enough to be used to decide whether differences between countries are likely to be statistically significant, but percentile confidence intervals tend to be conservative and are unreliable for small sample sizes, perhaps due to the ranking ties (Waltman & Schreiber, 2013), which make even the binomial distribution an approximation, as well as due to small sample sizes and proportions. New confidence interval formulae may be able to remedy this, however. The use of confidence intervals is particularly important given the imprecision of some of the indicators and a tendency for reports to include a range of indicators, perhaps focusing on those on which the target nation performs best. In this context it is possible that a nation performs particularly well on one of the percentile indicators by chance and therefore that the report would give a misleadingly optimistic assessment of the nation's research health. Nevertheless, this conclusion is based upon the assumption that the citation impact of every country's research outputs follows a lognormal distribution whereas, it is possible, for example, that a small country has one excellent research group and so could legitimately claim to excel in the top 1% indicator but not in the others. Hence, a detailed knowledge of a particular country may override statistical considerations.

Although the modelling was for comparisons between pairs of nations, the results are also applicable to other situations in which the impact of two groups of articles is compared, or when more groups of articles are compared pairwise, such as for research groups or departments within the same field, for states within the USA, or for geographic regions within the world (e.g., Europe vs. North America). The same is true for normalised data from the same field in multiple years, as long as the scale parameter for the lognormal distribution fit of the citation count data for the field does not change much between years.

All of the results assume that the samples are derived from distributions with the same scale parameter, even if their mean parameters differ. If the scale parameters are known to differ then the arithmetic mean or a top X% indicator may be preferred over the geometric mean if it is believed to be a better measure of central tendency in a particular context, for example because of the higher weighting it gives to highly cited articles. Conversely, the geometric mean may be an intrinsically better measure if the set of articles assessed cannot be adequately field normalised (Fleming & Wallace, 1986). Finally, the results are based upon models of a single document type and field, whereas international comparisons are typically based upon aggregated cross-field data. It is not clear whether this aggregation process affects the results, assuming that the different fields follow their own discretised lognormal distributions with different scale parameters.

# 9. Appendix: The ineffectiveness of standard statistical tests for the proportion in the top X%

The t-test cannot be used for a hypothesis test to assess whether the proportion of articles in the top X% most highly cited varies between two sets of articles, because the data is highly skewed. This is particularly evident for the top 1% and for small countries with a below average rate of attracting citations, because the sample mean (out of the 1000 iterations) of the proportions to be compared may be substantially smaller than 0.01, relative to a modest country sample size of as low as 25 (5% for a field of 500 publications). A log transformation cannot fully remove the skewing because in some cases the lowest value, zero, is the mode (for some parameter sets, over 80% of iterations of a sample have 0 articles in the top 1%).

It is also not possible to use one of the standard non-parametric tests that are used to compare means in similar case, the Mann-Whitney U test and the 2-sample Kolmogorov Smirnov test. This is because these tests do not directly test for differences in the means but test for differences in distributions. If two samples have different sizes but the same parameter sets then the proportion of articles in the top 1% will come from different distributions because the sample sizes are different. For example, a country publishing 25 articles out of 500 in a field and year could have 0, 1/25,2,25,...1 of their articles in the top 1% whereas a country publishing 75 articles in out of 500 in a field and year could have 0, 1/75, 2/75,....1 of their articles in the top 1% (ignoring ties). If these two countries were modelled by being sampled 1000 times from identical distributions then either of the two non-parametric tests could therefore generate very strong evidence to reject the null hypothesis that the proportion in the top 1% in the two samples were drawn from the same distribution. Table 3 shows parameters taken from one of the 1375 in Table 1, except that both samples were drawn from the same distribution with only the sample sizes being different. Using these parameters, the null hypothesis that the distributions of the two samples are the same was rejected by both the Mann-Whitney U test and 2-sample Kolmogorov-Smirnov test, at a significance level of 0.001, even though the sample means were almost identical (0.007943 and 0.007576). For example, the Mann-Whitney test (in R with the exact option) gave a $p$ value of $6 \times 10^{-18}$.

Intuitively, the reason why the two sets of proportions are radically different is that 80.5% of the 1000 copies of the smaller sample had no articles in the top 1% whereas only 53.5% of the copies of the larger sample had no articles in the top 1%. This naturally occurs because of the larger sample size but results in a randomly selected copy of sample 2 being more likely to have a lower proportion of articles in the top 1% than a randomly selected copy of sample 1. Thus the smaller sample size is more skewed than the larger sample size, despite having the same expected mean, and hence the Mann-Whitney U test and 2-sample Kolmogorov-Smirnov test can be expected to reject the null hypothesis that the two samples are taken from the same distribution.

**Table 3**. Lognormal distribution parameters for two samples of different sizes but otherwise identical properties.

| Statistic | Value |
|---|---|
| Overall mean parameter | 1 |
| Sample 1 mean parameter | 0.9 |
| Sample 2 mean parameter | 0.9 |
| Standard deviation parameter | 1 |
| Overall sample size | 500 |
| Sample 1 size | 75 |
| Sample 2 size | 25 |

The following concrete example illustrates the issue in a simplified version that excludes the possibilities of ties for presence in the top 1% (i.e., assuming that a country cannot have a fraction of an article in the top 1% due to ties). Suppose that country 1 and country 2 both operate for 1000 years with the unchanging parameters of sample 1 and sample 2 in Table 3, respectively, so that they have identically powerful underlying science systems but have different sizes. Randomly generating a set of articles for each country and each year, the data in Table 4 might occur. Although the arithmetic mean number of articles in the world's top 1% for the field is identical for both countries over 1000 years (0.00784), the rank sum of the first sample is much higher than the rank sum for the second sample, showing that for most years country 2 has fewer articles in the top 1% than country 2. This rank sum difference also causes a rejection of the null hypothesis in the Mann-Whitney U test with a calculated p value of 0.000. Note that since the values for country 2 are less than the values for country 1 for most years, despite their equal arithmetic means, a simple bootstrapping comparison would also reject the null hypothesis that the two samples (countries) are drawn from the same distribution.

**Table 4**. Results of modelling the countries with the properties in Table 3 for 1000 separate years. Both countries are randomly sampled from the same discretised lognormal distribution for each year but country 1 has 75 articles each year and country 2 has 25 articles each year. This is a simplified case that excludes the possibility of ties at the boundary of the top 1%.

| Country 1 no. of articles in the top 1% out of 75 | Country 1 prop. of articles in top 1% | Country 1 freq. (years) | Country 2 no. of articles in the top 1% out of 25 | Country 2 prop. of articles in top 1% | Country 2 freq. (years) | Average rank within the combined set | Country 1 rank sum | Country 2 rank sum |
|---|---|---|---|---|---|---|---|---|
| 0 | 0 | 534 | 0 | 0 | 815 | 675 | 360450 | 550125 |
| 1 | 0.0133 | 359 | | | | 1529 | 548911 | 0 |
| 2 | 0.0267 | 94 | | | | 1755.5 | 165017 | 0 |
| 3 | 0.04 | 11 | 1 | 0.04 | 174 | 1895 | 20845 | 329730 |
| 4 | 0.0533 | 2 | | | | 1988.5 | 3977 | 0 |
| 5 | 0.0667 | | | | | 1989.5 | 0 | 0 |
| 6 | 0.08 | | 2 | 0.08 | 11 | 1995 | 0 | 21945 |
| **Total** | | **1000** | | | **1000** | | **1099200** | **901800** |